%
%
%
\font\ninerm=cmr9
\font\eightrm=cmr8
\font\sixrm=cmr6
\font\ninei=cmmi9
\font\eighti=cmmi8
\font\sixi=cmmi6
\skewchar\ninei='177 \skewchar\eighti='177 \skewchar\sixi='177
\font\ninesy=cmsy9
\font\eightsy=cmsy8
\font\sixsy=cmsy6
\skewchar\ninesy='60 \skewchar\eightsy='60 \skewchar\sixsy='60

\font\ninebf=cmbx9
\font\eightbf=cmbx8
\font\sixbf=cmbx6
\font\ninett=cmtt9
\font\eighttt=cmtt8
\hyphenchar\tentt=-1 
\hyphenchar\ninett=-1
\hyphenchar\eighttt=-1
\font\ninesl=cmsl9
\font\eightsl=cmsl8
\font\nineit=cmti9
\font\eightit=cmti8
\newskip\ttglue
\def\tenpoint{\def\rm{\fam0\tenrm}%
  \textfont0=\tenrm \scriptfont0=\sevenrm \scriptscriptfont0=\fiverm
  \textfont1=\teni \scriptfont1=\seveni \scriptscriptfont1=\fivei
  \textfont2=\tensy \scriptfont2=\sevensy \scriptscriptfont2=\fivesy
  \textfont3=\tenex \scriptfont3=\tenex \scriptscriptfont3=\tenex
  \def\it{\fam\itfam\tenit}%
  \textfont\itfam=\tenit
  \def\sl{\fam\slfam\tensl}%
  \textfont\slfam=\tensl
  \def\bf{\fam\bffam\tenbf}%
  \textfont\bffam=\tenbf \scriptfont\bffam=\sevenbf
   \scriptscriptfont\bffam=\fivebf
  \def\tt{\fam\ttfam\tentt}%
  \textfont\ttfam=\tentt
  \tt \ttglue=.5em plus.25em minus.15em
  \normalbaselineskip=12pt
  \let\sc=\eightrm
  \let\big=\tenbig
  \setbox\strutbox=\hbox{\vrule height8.5pt depth3.5pt width0pt}%
  \normalbaselines\rm}
\def\ninepoint{\def\rm{\fam0\ninerm}%
  \textfont0=\ninerm \scriptfont0=\sixrm \scriptscriptfont0=\fiverm
  \textfont1=\ninei \scriptfont1=\sixi \scriptscriptfont1=\fivei
  \textfont2=\ninesy \scriptfont2=\sixsy \scriptscriptfont2=\fivesy
  \textfont3=\tenex \scriptfont3=\tenex \scriptscriptfont3=\tenex
  \def\it{\fam\itfam\nineit}%
  \textfont\itfam=\nineit
  \def\sl{\fam\slfam\ninesl}%
  \textfont\slfam=\ninesl
  \def\bf{\fam\bffam\ninebf}%
  \textfont\bffam=\ninebf \scriptfont\bffam=\sixbf
   \scriptscriptfont\bffam=\fivebf
  \def\tt{\fam\ttfam\ninett}%
  \textfont\ttfam=\ninett
  \tt \ttglue=.5em plus.25em minus.15em
  \normalbaselineskip=10pt 
  \let\sc=\sevenrm
  \let\big=\ninebig
  \setbox\strutbox=\hbox{\vrule height8pt depth3pt width0pt}%
  \normalbaselines\rm}
\def\eightpoint{\def\rm{\fam0\eightrm}%
  \textfont0=\eightrm \scriptfont0=\sixrm \scriptscriptfont0=\fiverm
  \textfont1=\eighti \scriptfont1=\sixi \scriptscriptfont1=\fivei
  \textfont2=\eightsy \scriptfont2=\sixsy \scriptscriptfont2=\fivesy
  \textfont3=\tenex \scriptfont3=\tenex \scriptscriptfont3=\tenex
  \def\it{\fam\itfam\eightit}%
  \textfont\itfam=\eightit
  \def\sl{\fam\slfam\eightsl}%
  \textfont\slfam=\eightsl
  \def\bf{\fam\bffam\eightbf}%
  \textfont\bffam=\eightbf \scriptfont\bffam=\sixbf
   \scriptscriptfont\bffam=\fivebf
  \def\tt{\fam\ttfam\eighttt}%
  \textfont\ttfam=\eighttt
  \tt \ttglue=.5em plus.25em minus.15em
  \normalbaselineskip=9pt
  \let\sc=\sixrm
  \let\big=\eightbig
  \setbox\strutbox=\hbox{\vrule height7pt depth2pt width0pt}%
  \normalbaselines\rm}
%
\def\headtype{\ninepoint}                 
\def\abstracttype{\ninepoint}             
\def\captiontype{\ninepoint}              
\def\footnotetype{\ninepoint}             
\def\refit{\it}                           
\font\chaptitle=cmr10 at 11pt             
\rm                                       

%
%
\parindent=0.25in                         
\parskip=0pt                              
\baselineskip=12pt                        
\hsize=4.25truein                         
\vsize=7.445truein                        
\hoffset=1in                              
\voffset=-0.5in                           

\newskip\sectionskipamount                
\newskip\aftermainskipamount              
\newskip\subsecskipamount                 
\newskip\firstpageskipamount              
\newskip\capskipamount                    
\newskip\ackskipamount                    
\sectionskipamount=0.2in plus 0.09in
\aftermainskipamount=6pt plus 6pt         
\subsecskipamount=0.1in plus 0.04in
\firstpageskipamount=3pc
\capskipamount=0.1in
\ackskipamount=0.15in
\def\sectionskip{\vskip\sectionskipamount}
\def\aftermainskip{\vskip\aftermainskipamount}
\def\subsecskip{\vskip\subsecskipamount} 
\def\firstpageskip{\vskip\firstpageskipamount}

%
%
\nopagenumbers                            
\newcount\firstpageno                     
\firstpageno=\pageno                      
\newcount\chapno                          

\def\rightheadline{\headtype\phantom{\folio}\hfil\runningtitletext\hfil\folio}
\def\leftheadline{\headtype\folio\hfil\runningauthortext\hfil\phantom{\folio}}
\headline={\ifnum\pageno=\firstpageno\hfil
           \else
              \ifdim\ht\topins=\vsize           
                 \ifdim\dp\topins=1sp \hfil     
                 \else
                     \ifodd\pageno\rightheadline\else\leftheadline\fi
                 \fi
              \else
                 \ifodd\pageno\rightheadline\else\leftheadline\fi
              \fi
           \fi}

\def\bottomnumber{\hss\tenrm[\folio]\hss}
\footline={\ifnum\pageno=\firstpageno\bottomnumber\else\hfil\fi}

%
%
%
%
\outer\def\mainsection#1
    {\vskip 0pt plus\smallskipamount\sectionskip
     \message{#1}\vbox{\noindent{\bf#1}}\nobreak\aftermainskip\noindent}
 
\outer\def\subsection#1
    {\vskip 0pt plus\smallskipamount\subsecskip
     \message{#1}\vbox{\noindent{\bf#1}}\nobreak\smallskip\nobreak\noindent}
 

\def\title#1{{\chaptitle\leftline{#1}}}
\def\name#1{\leftline{#1}}
\def\affiliation#1{\leftline{\it #1}}
\def\abstract#1{{\abstracttype \noindent #1 \smallskip\vskip .1in}}
\def\ref{\noindent \parshape2 0truein 4.25truein 0.25truein 4truein}
\def\caption{\noindent \captiontype
             \parshape=2 0truein 4.25truein .125truein 4.125truein}

\def\footnote#1{\edef\fspafac{\spacefactor\the\spacefactor}#1\fspafac
      \insert\footins\bgroup\footnotetype
      \interlinepenalty100 \let\par=\endgraf
        \leftskip=0pt \rightskip=0pt
        \splittopskip=10pt plus 1pt minus 1pt \floatingpenalty=20000
        \textindent{#1}\bgroup\strut\aftergroup\strut\egroup\let\next}
\skip\footins=12pt plus 2pt minus 4pt 
\dimen\footins=30pc 

%
%

\def\@{\spacefactor 1000}

\def\,{\pcomma} 
\def\pcomma{\relax\ifmmode\mskip\thinmuskip\else\thinspace\fi}

\def\oversim#1#2{\lower0.5ex\vbox{\baselineskip=0pt\lineskip=0.2ex
     \ialign{$\mathsurround=0pt #1\hfil##\hfil$\crcr#2\crcr\sim\crcr}}}
\def\simgt{\mathrel{\mathpalette\oversim>}}
\def\simlt{\mathrel{\mathpalette\oversim<}}

\def\runningtitletext{The Structure and Evolution of Molecular Clouds}
\def\runningauthortext{Williams, Blitz, \& McKee}

\def\kms    {${\rm km~s^{-1}}$}
\def\cc     {${\rm cm^{-3}}$}

\def\eee    {$^{-3}$}

\def\nh     {n_{\rm H}}

\def\nhtwo  {n_{\rm H_2}}

\def\thirteenco   {$^{13}{\rm CO}$}
\def\calm{{\cal M}}
\def\calt{{\cal T}}
\def\calw{{\cal W}}
\def\mbe{M_{\rm BE}}
\def\mcr{M_{\rm cr}}
\def\mphi{M_\Phi}
\def\pth{P_{\rm th}}
\def\vcl{{V_{\rm cl}}}
\def\frac#1#2{{#1\over#2}}

\null

\tolerance=10000
\firstpageskip

{\baselineskip=14pt
\title{THE STRUCTURE AND EVOLUTION OF MOLECULAR CLOUDS:}
\title{FROM CLUMPS TO CORES TO THE IMF}
}

\vskip .3truein
\name{JONATHAN P. WILLIAMS}
\affiliation{Harvard--Smithsonian Center for Astrophysics}
\vskip .2truein
\name{LEO BLITZ and CHRISTOPHER F. MCKEE}
\affiliation{University of California at Berkeley}
\vskip .3truein

\abstract{We review the progress that has been made in observing and
analyzing molecular cloud structure in recent years.
Structures are self-similar over a wide range of scales
with similar power law indices independent of the
star forming nature of a cloud. Comparison of structures at
parsec-scale resolution in a star forming and non-star forming cloud
show that the average densities in the former are higher but the
structural characteristics in each cloud are much the same.
In gravitationally bound regions of a cloud, however,
and at higher densities and resolution, the self-similar scaling
relationships break down and it is possible to observe
the first steps toward star formation.
High resolution observations of the dense individual star forming
cores within the clumps hold the key to an empirical
understanding of the origins of the stellar initial mass function.}

\mainsection{I.~~INTRODUCTION}

Molecular clouds are generally self-gravitating, magnetized,
turbulent, compressible fluids. The puzzle of how stars form 
from molecular clouds begins with
understanding the physics of such objects and how individual,
gravitationally unstable cores condense within them.

In this review, we describe advances in understanding that have been
made since the last {\it Protostars and Planets} meeting
(see the chapter by Blitz in particular, but also chapters by
Lada, Strom \& Myers, Elmegreen, McKee et al., and Heiles et al.)
The predominant point of view at that time was that the inhomogeneous
structure that had been observed even in the earliest complete maps of
molecular clouds (e.g., Kutner et al. 1977; Blitz \& Thaddeus 1980)
could be described as a set of discrete clumps (Blitz \& Stark 1986).
These clumps themselves contain dense cores which are the localized
sites of star formation within the cloud (Myers \& Benson 1983).

In the intervening years there have been a growing number of papers
that describe an alternative point of view: that clouds are scale-free
and their structure is best described as fractal
(e.g., Bazell \& D\'{e}sert 1988; Scalo 1990;
Falgarone, Phillips, \& Walker 1991).
In this picture, the hierarachy of cores within clumps within clouds
is simply an observational categorization of this self-similar structure.
Here we contrast, but also try to reconcile, these two descriptions
with a focus on the global questions that link star formation to
the ISM: what controls the efficiency and rate of star formation,
and what determines the shape of the stellar initial mass function (IMF)?

We begin by describing the large scale view of the molecular ISM,
and move to progressively smaller scales. The study of molecular
clouds is a broad topic and to stay close to the title of this meeting,
we concentrate on the structure and evolution of the star forming regions
in these clouds.

\mainsection{{I}{I}.~~THE LARGE SCALE VIEW}

Advances in millimeter-wave receiver technology have made it possible
to map molecular clouds rapidly at high sensitivity.
The noise in an observation is directly proportional to the system
temperature, and as receiver temperatures have decreased, it has become
feasible to map entire complexes -- degrees in angular size -- at
sub-arcminute resolution.
Focal plane arrays 
(arrays of receivers observing neighboring points
of the sky simultaneously) at the FCRAO 14~m and NRAO 12~m single dish
telescopes have increased the mapping speed by an order of magnitude.
Up to four receivers operating at different frequencies can be used
simultaneously at the IRAM 30~m telescope to observe the same position
on the sky.
The expansion of the millimeter-wave interferometers, IRAM, OVRO, and
BIMA has dramatically increased the imaging quality and sensitivity of
these instruments and has made high resolution observations ($<10''$)
of structure in molecular clouds considerably quicker and easier.

Most of the mass of the molecular ISM is in the form of giant molecular
clouds (GMCs), with masses $\sim 10^{5-6}~M_\odot$, diameters $\sim 50$~pc,
and average densities, $\langle\nhtwo\rangle\sim 10^2$~\cc\
(e.g., Blitz 1993). The sharp cutoff at the upper end of the cloud mass
distribution at $\sim 6\times 10^6~M_\odot$ (Williams \& McKee 1997)
indicates that cloud masses are limited by some physical process,
such as the tidal field of the Galaxy or
the disrupting effect of massive stars within them.

The FCRAO telescope has recently completed a sensitive, high resolution,
unbiased study of the CO emission in the outer Galaxy (Heyer et al. 1998).
There is no distance ambiguity in the outer Galaxy and much less
blending of emission, and therefore this survey allows a more detailed
investigation of the large scale structure of the ISM (see Figure~1)
than earlier surveys of the inner Galaxy by Dame et al. (1987) and
Sanders et al. (1985). 
Heyer et al. (1998) confirm that there are large regions with little
or no CO emission (Dame et al. 1986) and concur with earlier results
that these regions have been cleared of molecular gas by the intense
radiation fields and stellar winds from massive stars.
Further recent observations of the structural imprint of massive
star formation in individual star forming regions is discussed by
Patel et al. (1995), Carpenter, Snell, \& Shloerb (1995a),
and Heyer, Carpenter, \& Ladd (1996).
Carpenter, Snell, \& Shloerb (1995b) suggest that massive stars
can act to compress gas and create dense cores that give rise
to the next generation of star formation as originally proposed by
Elmegreen \& Lada (1977).

Heyer \& Terebey (1998) and Digel et al. (1996) show that CO emission
in the outer Galaxy is confined almost exclusively to the spiral arms,
confirming the earlier results of Cohen et al. (1980).
The former, most recent work shows that the ratio of emission in the arm
to interarm regions is greater than 28:1.
The absence of molecular gas in the interarm regions implies that
molecular clouds form from a compressed atomic medium, and have
lifetimes that are less than an arm crossing time $\sim 10^7$~yr.
These conclusions may not apply in the inner Galaxy where it has
been more difficult to isolate arm emission and calculate an
arm-to-interarm ratio. 
Solomon \& Rivolo (1989) have argued that about half the
CO in the inner Galaxy is in clouds that are not actively
forming stars, and that this gas is not concentrated
in spiral arms.
However, Loinard et al. (1996) find an arm-to-interarm
ratio $\sim 10:1$ in M31 in the south-west ``ring'' $\sim 9$~kpc
from the galaxy center which suggests that similar conclusions
about cloud formation and lifetimes can be drawn in that galaxy too.

There have been several recent studies of HI halos around molecular
clouds (e.g. Kuchar \& Bania 1993; Williams \& Maddalena 1996;
Moriarty-Schieven, Andersson \& Wannier 1997)
and a comprehensive survey is underway at the DRAO telescope
(Normandeau, Taylor, \& Dewdney 1997).
In order to form a GMC of mass $10^6~M_\odot$  out of an atomic
ISM of density $\langle\nh\rangle\sim 1$~\cc,
gas must be accumulated from a volume
$\sim 0.4$~kpc in diameter.
Since such a large region encompasses many atomic clouds, the
density inhomogeneities in molecular clouds may simply reflect the initial
non-uniform conditions of their formation rather than the first step in the
fragmentation/condensation process that leads to the creation of stars
(Elmegreen 1985).
To distinguish between the remnants of the formation of a cloud and the
first steps toward star formation, it is necessary to analyze and
compare structures in a number of different clouds.

\mainsection{{I}{I}{I}.~~CLOUD STRUCTURE AND SELF-SIMILARITY}
\subsection{A. A categorization of molecular cloud structure}

	Before we discuss the analysis of cloud structure, we first define
an operational categorization into clouds, clumps, and cores.
This categorization is not inconsistent with the 
fractal models for cloud structure that are discussed in \S III.E,
although we argue that gravity introduces scales that limit
the range of validity of the fractal description.

	Molecular clouds are regions in which the gas is primarily molecular.
Almost all known molecular clouds in the Galaxy are detectable in CO.
Giant molecular clouds have masses $\simgt 10^4~M_\odot$,
are generally gravitationally bound, and may contain several sites
of star formation. However, there also exist
small molecular clouds with masses $\simlt 10^2~M_\odot$,
such as the unbound high latitude clouds discovered by
Blitz, Magnani, \& Mundy (1984),
and the small, gravitationally bound molecular clouds in the
Galactic plane cataloged by Clemens \& Barvainis (1988).
A small number of low mass stars are observed to form in some
of these clouds but the contribution to their total star formation rate
in the Galaxy is negligible (Magnani et al. 1995).

	Clumps are coherent regions in $l-b-v$ space, generally identified
from spectral line maps of molecular emission.
Star-forming clumps are the massive clumps out of which stellar clusters form.
Although most clusters are unbound, the gas out of which
they form is bound (Williams, Blitz, \& Stark 1995).
Cores are regions out of which single stars (or multiple systems such as
binaries) form and are necessarily gravitationally bound. 
Not all material that goes into forming a star must come from
the core; some may be accreted from the surrounding clump
or cloud as the protostar moves through it (Bonnell et al. 1997).

	We have discussed this categorization to describe observations
in a uniform manner and to provide a clear link to the processes of
star formation. In the following sections we adopt this classification
and discuss the relationship between the structure 
of molecular clouds and
their evolution toward star formation.

\subsection{B. The virial theorem for molecular clouds}

	The condition for a molecular cloud or clump within
it to be gravitationally bound can be inferred
from the virial theorem, which may be written
$$
\frac 12 \ddot I=2(\calt-\calt_0)+\calm+\calw,
\eqno(1)
$$
where $I$ is the moment of inertia,
$\calt$ is the total kinetic energy (including
thermal), $\calm$ is the net magnetic energy, and
$\calw$ is the gravitational energy (see McKee et al.
1993 for more details).  The moment of inertia term
is usually neglected, but it may be significant for
a turbulent cloud.  In contrast to the terms on the RHS
of the equation, it can be of either sign, and
as a result its effects can be averaged out either by
applying the virial theorem to an ensemble of clouds or clumps
or by averaging over a time long compared with the dynamical
time (assuming the cloud lives that long.)  The kinetic energy term,
$$
\calt=\int_\vcl\left(\frac 32 \pth+\frac 12\rho v^2\right)\;
	dV\equiv \frac 32 \bar P \vcl,
\eqno(2)
$$
includes both the thermal and non-thermal pressure inside
the cloud.  The surface term can be expressed
as $\calt_0=\frac 32 P_0\vcl$, where $P_0$ is
about equal to the total thermal plus
nonthermal pressure in the ambient medium
(McKee and Zweibel 1992).  Finally, the
gravitational term can be written as (Bertoldi and McKee 1992)
$$
\calw\equiv -3P_G\vcl,
\eqno(3)
$$
where the ``gravitational pressure"
$P_G$ is the mean weight of the material in the cloud.
With these results, the steady--state, or time--averaged,
virial theorem becomes
$$
\bar P=P_0+P_G\left(1-\frac{\calm}{\vert\calw\vert}\right).
\eqno(4)
$$
In this form, the virial theorem has an immediate intuitive
meaning: the mean pressure inside the cloud is the surface
pressure plus the weight of the material inside the cloud,
reduced by the magnetic stresses.

	In the absence of an external gravitational field,
$\calw$ is the gravitational energy of the cloud,
$$
\calw=-\frac 35 a\left(\frac{GM^2}{R}\right),
\eqno(5)
$$
where $a$ is a numerical factor of order unity that
has been evaluated by Bertoldi and McKee (1992).
The gravitational pressure is then
$$
P_G=\left(\frac{3\pi a}{20}\right) G\Sigma^2
	\rightarrow 5540\bar A_V^2 ~~~~{\rm K\ cm^{-3}},
\eqno(6)
$$
where $\Sigma$ is the mean surface density of the cloud
and $\bar A_V$ is the corresponding visual extinction.
          The numerical evaluation is for a spherical
cloud with a $1/r$ density profile.

	Magnetic fields play a crucial role in the 
structure and evolution of molecular clouds.
For poloidal fields,
the relative importance of gravity and magnetic field is determined 
by the ratio of the mass to the {\it magnetic
critical mass},
$$
\mphi=\frac{c_\Phi \Phi}{G^{1/2}}.
\eqno(7)
$$
The value of the numerical factor $c_\Phi$ depends
on the distribution of mass to flux.
If the mass--to--flux ratio is constant,
then $c_\Phi=1/2\pi$ (Nakano \& Nakamura 1978).  
In this case, the ratio of the magnetic force
to the gravitational force is (in our notation)
$(\mphi /M)^2$ (Shu \& Li 1997), which is invariant 
so long as the magnetic
flux is frozen to the matter.  For $M<\mphi$, the cloud
is said to be {\it magnetically subcritical}: such a cloud
can never undergo gravitational collapse (so long as
flux freezing holds) since the magnetic force always
exceeds the gravitational force.  Conversely,
if $M>\mphi$, the cloud is {\it magnetically supercritical},
and magnetic fields cannot prevent gravitational collapse.
Shu \& Li (1997) have presented a general analysis of the forces
in magnetized disks with a constant mass--to--flux ratio
(which they term ``isopedic" disks); their analysis
applies even if the disks are non-axisymmetric and time-dependent.
The distinction between magnetically subcritical and
supercritical disks holds for non--isopedic disks
as well, although the value of $c_\Phi$ may differ;
for example, for a mass--to--flux
distribution corresponding to a uniform field threading
a uniform, spherical cloud, Tomisaka et al. (1988) find
$c_\Phi\simeq 0.12$.  

	Toroidal fields can provide a confining force,
thereby reducing the magnetic critical mass (Tomisaka 1991;
Fiege \& Pudritz 1998).  However, the ratio of the toroidal
field to the poloidal field cannot become too large without
engendering instabilities (e.g., Jackson 1975).

	If the cloud is supported by thermal
and nonthermal pressure in addition to the magnetic stresses,
then the maximum stable mass is the critical mass $\mcr$,
and the cloud is subcritical for $M<\mcr$ and supercritical
for $M>\mcr$.  For an isothermal cloud, the critical mass is given by
$\mcr\simeq \mphi+\mbe$, where 
the Bonnor-Ebert mass $\mbe=1.18c_s^4/(G^3P_0)^{1/2}$
is the largest gravitationally stable mass at an exterior
pressure $P_0$ for a nonmagnetic sphere (McKee 1989).

	What do observations say about the importance
of magnetic fields in clouds?  The most detailed study
is that of the cloud B1, for which Crutcher et al. (1994)
found that the inner envelope was marginally magnetically
subcritical, whereas the densest region was somewhat
supercritical.  The observational results were shown to
be in good agreement with a numerical model that, however,
did not include the observed nonthermal motions.
Crutcher (1999) has summarized the observations
of magnetic fields in a number of clouds,
and finds that they are magnetically
supercritical; in this sample, which tends
to focus on the central regions of clouds,
there is no clear case
in which a cloud is magnetically subcritical.
If the magnetic field makes an angle $\theta$
with respect to the line of sight, then the
observed field is smaller than the true value
by a factor $\cos\theta$; on average, this
is a factor 1/2.  After allowing for this, he
finds that on average $\langle M/M_\Phi\rangle\simeq 2.5$.
If the clouds are flattened along the field lines,
then the observed area is smaller than the true
area by factor $\cos\theta$ as well, so that
$M/M_\Phi\propto \cos^2\theta$; on average, this
is a factor 1/3.  However, since clouds are observed
to have substantial motions, they are unlikely
to be highly flattened along field lines,
so Crutcher concludes that $\langle M/M_\Phi\rangle\simeq 2$.  The idea
that GMCs are magnetically supercritical with
$M/M_\Phi\simeq 2$ was
suggested theoretically by McKee (1989);
Bertoldi \& McKee (1992) extended this argument
to star forming clumps,
and Nakano (1998) argued that cores are magnetically
supercritical.  Crutcher (1999) also finds that the
Alfv\'en Mach number of the motions,
$m_A=\sigma\surd 3/v_A$, is about unity, as inferred
previously by others on the basis of less
complete data (e.g., Myers \& Goodman 1988).

	With these results in hand, we can now address the
issue of whether molecular clouds and their constituents
are gravitationally bound.  The virial
theorem (eq. 4) enables us to write the total energy 
$E=\calt+\calm+\calw$ as
$$
E=\frac 32 \left[P_0-P_G\left(1-\frac{\calm}{\vert\calw\vert}
	\right)\right]\vcl.
\eqno(8)
$$
In the absence of a magnetic field, the condition that the cloud
be bound (i.e., $E<0$) is simply $P_G>P_0$.  We shall use this criterion
even for magnetized clouds, bearing in mind that using
the total ambient gas pressure for $P_0$ is an overestimate
and that our analysis is approximate because we have 
used the time-averaged virial theorem.

	For GMCs, the surface pressure is that of the ambient interstellar
medium.  In the solar vicinity, the total interstellar pressure,
which balances the weight of the ISM,
is about $2.8\times 10^4$ K cm\eee\ (Boulares and Cox 1990).
Of this, about $0.7\times 10^4$ K cm\eee\ is due to cosmic
rays; since they pervade both the ISM and a molecular
cloud, they do not contribute to the support of a cloud
and may be neglected.  The magnetic pressure is about
$0.3\times 10^4$ K cm\eee\ (Heiles 1996), leaving
$P_0\simeq 1.8\times 10^4$ K cm\eee\ as the total ambient gas pressure.

	What is the minimum value of $P_G$ for a molecular
cloud?  According to van Dishoeck and Black (1988),
molecular clouds exposed to the local interstellar
radiation field have a layer of C$^+$ and C$^0$ corresponding
to a visual extinction of 0.7 mag.
If we require at least 1/3 of the carbon along
a line of sight through a cloud to be in the form of CO
in order for it to be considered ``molecular",
then the total visual extinction must be 
$\bar A_V>2$ mag (allowing for a shielding layer
on both sides).  According to equation (6), this
gives $P_G\simgt 2\times 10^4$ K cm\eee\ $\sim P_0$,
verifying that molecular clouds as observed in CO 
are at least marginally bound (e.g. Larson 1981).
Note that if we defined molecular clouds as having
a significant fraction of H$_2$ rather than CO,
the minimum column density required would be substantially less
and the clouds might not be bound.  Furthermore,
the conclusion that CO clouds are bound depends both
on the interstellar pressure and on the strength of the
FUV radiation field, and CO clouds may not be bound everywhere
in the Galaxy or in other galaxies (Elmegreen 1993b).

	GMCs in the solar neighborhood
typically have mean extinctions significantly
greater than 2 mag, and as a result $P_G$ is
generally significantly greater than $P_0$.
For GMCs in the solar neighborhood,  $P_G\sim 2\times 10^5$ K cm\eee,
an order of magnitude greater than $P_0$
(Blitz 1991; Bertoldi \& McKee 1992; Williams et al. 1995).
Thus if GMCs are dynamically stable entities (the
crossing time for a GMC is about 10$^7$ y, smaller than the expected
lifetime -- see Blitz \& Shu 1980),
then GMCs must be self-gravitating.
In the inner galaxy, where $P_0$ is expected to be
greater, the typical GMC linewidths also appear to be somewhat
greater than those found locally (Sanders, Scoville, \& Solomon
1985), and thus $P_G$ is still comfortably greater than $P_0$. 

	For clumps within GMCs, the surface pressure is just
the mean pressure inside the GMC, $P_G($GMC$)\propto \Sigma^2$(GMC),
so the virial theorem becomes $\bar P$(clump)$\propto
\Sigma^2$(GMC)$+\Sigma^2$(clump).  Most clumps
observed in $^{13}$CO have $\Sigma$(clump)$\ll
\Sigma$(GMC) and are therefore confined by
pressure rather than gravity; on the other hand,
much of the mass is in massive star-forming clumps
that have $\Sigma$(clump)$\simgt
\Sigma$(GMC) and are therefore gravitationally bound
(Bertoldi and McKee 1992).

\subsection{C. Structure analysis techniques}

	Molecular cloud structure can be mapped via radio spectroscopy of
molecular lines (e.g., Bally et al. 1987),
continuum emission from dust (e.g., Wood, Myers, \& Daugherty 1994),
or stellar absorption by dust (Lada et al. 1994).
The first gives kinematical as well as spatial information and
results in a three dimensional cube of data, whereas the latter two
result in two dimensional datasets.
Many different techniques have been developed to analyze these data
which we discuss briefly here.

	Stutzki \& G\"{u}sten (1990) and Williams, de Geus, \& Blitz (1994)
use the most direct approach and decompose the data into
a set of discrete clumps,
the first based on recursive tri-axial gaussian fits,
and the latter by identifying peaks of emission and
then tracing contours to lower levels.
The resulting clumps can be considered to be the ``building blocks''
of the cloud and may be analyzed in any number of ways to
determine a size-linewidth relation, mass spectrum, and variations
in cloud conditions as a function of position (Williams et al. 1995).
There are caveats associated with each method of clump deconvolution,
however. Since the structures in a spectral line map of a molecular cloud
are not, in general, gaussian, the recursive fitting method of
Stutzki \& G\"{u}sten (1990) will tend to find and
subsequently fit residuals around each clump, which
results in a mass spectrum that is steeper than
the true distribution. On the other hand, the contour tracing
method of Williams et al. (1994) has a tendency to blend small
features with larger structures and results in a mass spectrum
that is flatter than the true distribution.

	Heyer \& Schloerb (1997) use
principal component analysis to identify differences
in line profiles over a map. A series of eigenvectors and
eigenimages are created which identify ever smaller velocity
fluctuations and their spatial distribution, resulting in the
determination of a size-linewidth relation.
Langer, Wilson, \& Anderson (1993) use Laplacian pyramid transforms
(a generalization of the Fourier transform) to measure the power on
different size scales in a map; as an application, they
determine the mass spectrum in the B5 molecular cloud.
Recently, Stutzki et al. (1998) have described a closely related
Allan-variance technique to characterize the fractal structure of
2-dimensional maps.
Houlahan \& Scalo (1992) define an algorithm that constructs
a structure tree for a map; this retains the spatial relation
of the individual components within the map but loses information
regarding their shapes and sizes. It is most useful for displaying
and ordering the hierarchical nature of the structures in a cloud.

	Adams (1992) discusses a topological approach to quantify the
difference between maps. Various ``output functions''
(e.g., distribution of density, volume, and number of components
as a function of column density; see Wiseman \& Adams 1994)
are calculated for each cloud dataset and a suitably defined
metric is used to determine the distance between these
functions and therefore to quantify how similar clouds are,
or to rank a set of clouds.

	A completely different technique was pioneered by Lada et al. (1994).
They determine a dust column density in the dark cloud IC\,5146
by star counts in the near-infrared and mapped cloud structure over a
much greater dynamic range ($A_V=0-32$~mag) than a single spectral line map.

	The most striking result of applying these various analysis tools
to molecular cloud datasets is the identification of self-similar
structures characterized by power law relationships between,
most famously, the size and linewidth of features (Larson 1981),
and the number of objects of a given mass (e.g., Loren 1989).
Indeed, mass spectra are observed to follow a power law with
nearly the same exponent, $x=0.6-0.8$, where $dN/d\ln M \propto M^{-x}$
from clouds with masses up to $10^5~M_\odot$ in the outer Galaxy
to features in nearby high-latitude clouds with masses
as small as $10^{-4}~M_\odot$
(Heyer \& Terebey 1998; Kramer et al. 1998a; Heithausen et al. 1998).
Since a power law does not have a characteristic scale, the
implication is that clouds and their internal structure are scale-free.
This is a powerful motivation for a fractal description of the
molecular ISM (Falgarone et al. 1991, Elmegreen 1997a).
On the other hand, molecular cloud maps do have clearly identifiable
features, especially in spectral line maps when a velocity axis can be
used to separate kinematically distinct features along a line of sight
(Blitz 1993).
These features are commonly called clumps,
but there are also filaments (e.g., Nagahama et al. 1998),
and rings, cavities, and shells (e.g., Carpenter et al. 1995a).

\subsection{D. Clumps}

Clump decomposition methods such as those described above by
Stutzki \& G\"{u}sten (1990) and Williams et al. (1994)
can be readily visualized and have an appealing simplicity.
In addition, as for all automated techniques, these algorithms
offer an unbiased way to analyze datasets, and are still a valid
and useful tool for cloud comparisons even if one does not subscribe
to the notion of clumps within clouds as a physical reality (Scalo 1990).

In a comparative study of two clouds, Williams et al. (1994)
searched for differences in cloud structure between a star forming
and non-star forming GMC.
The datasets they analyzed were maps of \thirteenco(1--0) emission
with similar spatial (0.7~pc) and velocity resolution (0.68~\kms)
but the two clouds, although of similar mass $\sim 10^5~M_\odot$,
have very different levels of star formation activity. The first, the
Rosette molecular cloud, is associated with an HII region
powered by a cluster of 17 O and B stars and also contains a number
of bright infrared sources from ongoing star formation deeper
within the cloud (Cox, Deharveng, \& Leene 1990).
The second cloud, G216-2.5, originally discovered by
Maddalena \& Thaddeus (1985),
contains no IRAS sources from sites of embedded star formation
and has an exceptionally low far-infrared luminosity to mass ratio 
(Blitz 1990),
$L_{\rm IR}/M_{\rm cloud}<0.07~L_\odot/M_\odot$, compared to
more typical values of order unity (see Williams \& Blitz 1998).

Almost 100 clumps were cataloged in each cloud, and sizes, linewidths,
and masses were calculated for each. These basic quantities were found to be
related by power laws with the same index for the two clouds,
but with different offsets (Figure~2) in the sense that for a given mass,
clumps in the non-star forming cloud are larger, and have greater linewidths
than in the star forming cloud.
The similarity of the power law indices suggests that,
on these scales, $\sim$~few~pc,
and at the low average densities, $\langle\nhtwo\rangle\sim 300$~\cc,
of the observed clumps, the principal difference between the star forming
and non-star forming cloud is the change of scale rather than the
collective nature of the structures in each cloud.

Figure~2 shows that the kinetic energy of each clump in
G216-2.5 exceeds its gravitational potential energy, and
therefore no clump in the cloud is bound (although the
cloud as a whole is bound).
On the other hand, Williams et al. (1995) show that,
for the Rosette molecular cloud, stars formation occurs
only in the gravitationally bound clumps in the cloud.
Therefore, the lack of bound clumps in G216-2.5 may explain why
there is no star formation currently taking place within it.

Even in the Rosette cloud, most clumps are not gravitationally bound.
These unbound clumps have similar density profiles, $n(r)\propto 1/r^2$,
as the bound clumps (Williams et al. 1995), but contain relatively
little dense gas as traced by CO(3--2) or CS(2--1)
(Williams \& Blitz 1998).
The unbound clumps are ``pressure confined" in that their internal
kinetic pressure, which is primarily turbulent, is comparable to the
mean pressure of the ambient GMC (Blitz 1991; Bertoldi \& McKee 1992).
Simulations suggest that these clumps are transient
structures (see chapter by Ostriker et al.)

The nature of the interclump
medium remains unclear: Blitz \& Stark (1986) postulated that
it is low density molecular gas, but this has been questioned by
Schneider et al. (1996). Blitz (1990, 1993) showed that atomic
envelopes around molecular clouds are quite common and that the
atomic gas plausibly pervades the interclump medium.
Williams et al. (1995) show that the
HI associated with the molecular gas in the Rosette has about the same
turbulent pressure, and suggest that it could serve as the confining medium.

	Cloud, clump and core density profiles are reflections of the
physics that shape their evolution, but the density profiles of clouds
and clumps have received scant attention.  For clouds, which often
are quite amorphous without a clear central peak, the density profile
is often difficult to define observationally.  For clumps, Williams
et al. (1995) showed that surface density profiles of pressure bound,
gravitationally bound, and star-forming clumps all have similar power
law indices close to 1.  Formally, the
fits range from -0.8 to -1.2, but these differences do not appear to
be significant.  For a spherical cloud of infinite extent, $\Sigma(r)\propto
r^{-1}$ implies $\rho(r) \propto r^{-2}$, suggesting that the
(turbulent) pressure support is spatially constant. However, McLaughlin \&
Pudritz (1996) argued that for finite spheres, the volume density
distribution can be considerably flatter than that inferred for
infinite clumps. Density distributions inferred from observations
also require consideration of beam-convolution effects.  It is
nevertheless astonishing that both strongly self-gravitating clumps
and those bound by external pressure have such similar, perhaps
identical density distributions.  Why this should be so is unclear.

\subsection{E. Fractal Structures}

An alternate description of the ISM is based on fractals.
High spatial dynamic range observations of molecular
clouds, whether by millimeter spectroscopy (e.g., Falgarone et al. 1998),
IRAS (Bazell \& D\'{e}sert 1988), or using the Hubble Space Telescope
(e.g., O'Dell \& Wong 1996; Hester et al. 1996)
show exceedingly complex patterns that appear to defy a
simple description in terms of clouds, clumps and cores;
Scalo (1990) has argued that such loaded names arose from
lower dynamic range observations and a general human tendency to
categorize continuous forms into discrete units.

As we have discussed above, it seems that however one analyzes
a molecular cloud dataset, one finds self-similar structures.
Moreover, the highly supersonic linewidths that are observed in
molecular clouds probably imply turbulent motions
(see discussion in Falgarone \& Phillips 1990), for which one
would naturally expect a fractal structure (Mandelbrot 1982).

The fractal dimension of a cloud boundary, $D$,
can be determined from the perimeter-area relation of a map,
$P\propto A^{D/2}$. Many studies of the molecular ISM
find a similar dimension, $D\simeq 1.4$ (Falgarone et al. 1991 and
references therein). In the absence of noise, $D>1$ demonstrates that cloud
boundaries are fractal. That $D$ is invariant from cloud to cloud
(star-forming or quiescent, gravitationally bound or not)
is perhaps related to the similarity in the
mass spectrum index in many different molecular clouds
(Kramer et al. 1998a; Heithausen et al. 1998).
Fractal models have been used to explain both the observed mass spectrum of
structures (Elmegreen \& Falgarone 1996) and the stellar IMF (Elmegreen 1997b).

Probability
density functions (PDFs) may be used to describe the distribution of
physical quantities (such as density and velocity) in a region of space
without resorting to concepts of discrete objects
such as clouds, clumps and cores.
The density PDF is readily determined from simulations, and can
be semi-analytically modeled (Vazquez-Semadeni 1998)
but is very difficult to determine from observations because
projection and excitation effects result in maps of
column density integrated over a limited range of volume density.
We discuss column density PDFs in \S III.F.

	Velocities, albeit also projected, can be much more easily measured.
Falgarone \& Phillips (1990) show that the PDF of the velocity field
can be determined from high signal-to-noise observations of a single
line profile. The low-level, broad line wings that are observed in
non-star forming regions show that the probability of rare, high-velocity
motions in the gas are greater that predicted by a Normal (gaussian)
probability distribution. This {\it intermittent} behavior is expected
in a turbulent medium, and their detailed analysis shows that the
deviations from the predictions for Kolmogorov turbulence are small
despite the fact that the basic assumptions of
Kolmogorov turbulence, such as incompressibility and $B=0$, are not satisfied.
Miesch \& Scalo (1995) calculate velocity centroid
PDFs from \thirteenco\ observations of star-forming regions and
also report non-gaussian behaviour. Lis et al. (1996) compare their
results with a similar analysis applied to simulations of compressible
turbulence; such work may be a promising avenue for exploring the
role of turbulence in molecular clouds.

A key project at the IRAM 30~m telescope has been the investigation
of the small scale structure in pre-star forming regions
(Falgarone et al. 1998). Their study of many high resolution,
high spatial dynamic range maps in a number of
different tracers and transitions shows that the line core emission
has a rather smooth and extended spatial distribution,
but the line wing emission has a more filamentary distribution.
These filaments also show very large local velocity gradients,
$\simgt 10$~\kms\,pc$^{-1}$,
and have the greatest amount of small scale structure.

Falgarone et al. also address a long standing problem concerning
the extreme smoothness of observed line profiles. Isotopic ratios indicate
that CO lines, especially at line center, are very optically thick,
yet line profiles are generally neither flat topped nor self-reversed. 
Moreover, line profiles do not break up into separate components at high
angular and spectral resolution.
This has traditionally presented a problem for the clump-based
picture of molecular clouds (Martin, Sanders, \& Hills 1984).
Falgarone et al. interpret their results in terms of a
macroturbulent model: emission arises from a large number of
cells with size $\simlt 200$~AU and densities
ranging from $\nhtwo\sim 10^3$~\cc\ for line-wing emission,
to $\nhtwo\sim 10^5$~\cc\ for line-core emission.
They find that an anti-correlation between linewidth and intensity
and speculate that as the turbulent motions dissipate, there is
increased radiative coupling between cells.
Macroturbulent models have also been explored by
Wolfire, Hollenbach, \& Tielens (1993)
who include the effects of photodissociation,
and Tauber (1996) who includes an interclump medium but finds
no evidence for small scale structures ($\simlt 0.5$~pc) in Orion A.

\subsection{F. Departures from self-similarity}

The universal self-similarity that is observed in all types of cloud,
over a wide range in mass and star forming activity is remarkable,
but a consequence of this universality is that it does not differentiate
between clouds with different rates of star formation (or those that are
not forming stars at all) and therefore it cannot be expected to explain
the detailed processes by which a star forms.
Star formation must be preceeded
by a departure from structural self-similarity.

	An upper limit to the range over which self-similar
scaling laws apply is set by the generalization of the
Bonnor-Ebert mass to non-thermal linewidths.
From an analysis of the clumps in Ophiuchus,
Orion B, the Rosette, and Cep OB 3, Bertoldi \& McKee (1992) found that 
$M_{\rm BE}=1.18\sigma^4/(G^3P)^{1/2}$ is about constant for
all the pressure-confined clouds in each GMC. Here, $\sigma$ is the
total (thermal $+$ non-thermal) velocity dispersion, For all but
Cep OB 3, for which the data are of low resolution,
star formation was confined to clumps with $M\simgt M_{\rm BE}$,
and essentially all the clumps with $M>M_{\rm BE}$ are forming
stars.  The maximum mass of a clump in these clouds
was in the range $(1-10) M_{\rm BE}$
(for Cep OB 3, this statement applies only to the limited
part of the cloud for which data were available).

	On the small scale, there have long been suggestions
that the thermal Bonnor-Ebert mass gives a scale 
that determines the characteristic mass of stars (Larson 1985).
In order to determine whether such a scale is important in molecular clouds,
Blitz \& Williams (1997) 
examined how the structural properties of a large scale, high resolution
\thirteenco\ map of the Taurus molecular cloud obtained
by Mizuno et al. (1995) varied as the resolution was degraded by
an order of magnitude. In their work, they use the temperature histogram
of the dataset to compare the cloud properties as a function of resolution.
This is the most basic statistic and requires minimal interpretation
of the data. In Figure~3 we show the column density PDF,
which is proportional to the integrated intensity histogram
for optically thin emission, for this Taurus dataset at
two resolutions and four other \thirteenco\ maps of molecular clouds.

	To compare the different cloud PDFs, Figure~3 shows column
densities that have been normalized by the peak,
$N_{\rm peak}$, of each map.
Each PDF has also been truncated at $N/N_{\rm peak}\simeq 0.15-0.25$
to show only those points with high signal-to-noise.
Within the Poisson errors (not shown for clarity), the PDFs of
the different clouds are all the same, except for the higher
resolution Taurus PDF for which there is a lower relative
probability of having lines of sight with $N/N_{\rm peak}\simgt 0.7$.
The low intensities of the \thirteenco\ emission imply that the optical
depth is small along all lines of sight in the map, and is not responsible
for this effect. Rather, from examination of the integrated intensity maps,
Blitz \& Williams (1997) show that this is due to a steepening of the
column density profiles at small size scales (see also Abergel et al. 1994).

	There are two immediate implications from Figure~3.
First, the common exponential shape for the column density PDF
is another manifestation of the self similar nature of cloud structure.
It is a simple quantity to calculate and may provide a quick and useful
test of the fidelity of cloud simulations.

	Second, since the behavior of the Taurus dataset changes as it is
smoothed, it cannot be described by a single fractal dimension over
all scales represented in the map (see Williams 1998 for a demonstration
that the intensity histogram of a fractal cloud is resolution invariant
as one would expect).
There is other evidence for departures from self-similarity at 
similar size
scales. Goodman et al. (1998) examine in detail the nature of the
size-linewidth relation in dense cores as linewidths approach a constant,
slightly greater than thermal, value in a central ``coherent'' region
$\sim 0.1$~pc diameter (Myers 1983).
Also, Larson (1995) finds that the two-point angular correlation function
of T Tauri stars in Taurus departs from a single power law at a size scale
of 0.04~pc (see also Simon 1997).

For gas of density $\nhtwo\sim 10^3$~\cc, these size scales
correspond to masses of order $\sim 1~M_\odot$, close to the
thermal Bonnor-Ebert mass at a temperature $T=10$~K. 
It must be emphasized that the
above evidence for characteristic scales comes from studies of
gravitationally bound, star forming regions:
self-similarity in unbound clouds continues to much smaller scales.
Figure~3 shows that the column density PDF
of the unbound, high latitude cloud MBM12 is identical, at a
resolution of 0.03~pc, to the other lower resolution PDFs
of star forming GMCs. Similarly, the mass spectra of other
high latitude clouds follow power laws,
$dN/d\ln M \propto M^{-x}$ with $x\simeq 0.6-0.8$,
down to extremely low masses, $M\simeq 10^{-4}~M_\odot$
(Kramer et al. 1998a; Heithausen et al. 1998).
It appears to be the action of gravity that creates the
observed departures from self-similarity.

\mainsection{{I}{V}.~~THE CONNECTION BETWEEN CLOUD STRUCTURE 
AND STAR FORMATION}

\subsection{A. Star-forming clumps}

Star-forming clumps are bound regions in a molecular cloud that
form stellar clusters.
Since most stars form in clusters, questions of star formation
efficiency and rate are tied into the efficiency and rate of 
formation of star-forming clumps, 
and the IMF is related to the fragmentation of such clumps into
individual star forming cores.

The median column density of molecular gas in the outer Galaxy CO survey by
Heyer et al. (1998) is only $N({\rm H_2})\simeq 2\times 10^{21}$~cm$^{-2}$,
and most of the mass of a molecular cloud is contained within the low
column density lines of sight, $A_V\simlt 2$
(Carpenter et al. 1995a; Heyer et al. 1996).
Such gas is ionized predominantly by the interstellar
far ultraviolet radiation field ($h\nu<13.6$~eV).
McKee (1989) showed that this is true throughout the Galaxy:
most molecular gas is photoionized and therefore has a higher level
of ionization than that due to cosmic rays alone.  Since the rate of
low--mass star formation may be governed by ambipolar diffusion,
which in turn is determined by the ionization (e.g., Shu et al. 1987),
low--mass star formation is ``photoionization--regulated":
most stars form in regions shielded from photoionization
by column densities of dust corresponding to $A_V\simgt 3-4$.
This idea has been applied to the formation of star
clusters by Bertoldi \& McKee (1996).
It naturally accounts for the low average
rate of star formation in the Galaxy,
since only about 10\% of the mass of a typical GMC is
sufficiently shielded to have active star formation,
and the ambipolar diffusion timescale is about 10
times the free fall time. Observations of the ISM in the
low metallicity Magellanic Clouds have verified the prediction that
molecular clouds have about the same dust column densities, and therefore
higher gas column densities, as Galactic molecular clouds (Pak et al. 1998).
Li, Evans, \& Lada (1997) have tested photoionization--regulated
star formation by searching for evidence of recent star
formation in low extinction regions of L1630; they
found no evidence for such star formation.
On the other hand, Strom et al. (1993) did find
evidence for distributed star formation in
L1641; possible reasons for the discrepancy are discussed by Li et al.
In addition, Nakano (1998) has questioned whether ambipolar diffusion
plays any important role in low mass star formation.
He has correctly pointed out that cores are magnetically
supercritical, so that ambipolar diffusion is unimportant there,
but he has not addressed the issue of whether ambipolar diffusion
was important prior to the formation of the cores.  
An observational determination of the role of ambipolar diffusion
in low mass star formation remains a challenge for the future.

\subsection{B. Cores}

The core that forms an individual star (or multiple stellar system) is
the final stage of cloud fragmentation. Cores have typical average densities
$\nhtwo\sim 10^5$~\cc\ and can be observed in high excitation lines
or transitions of molecules with large dipole moments (Benson \& Myers 1989),
or via dust continuum emission at millimeter and sub-millimeter
wavelengths (Kramer et al. 1998b).

Because of their high densities, the surface filling fraction of
cores is low, even in cluster forming environments.
Therefore searches for cores have generally followed signs of
star formation activity, e.g. IRAS emission, outflows, etc.
and there have been few unbiased searches (e.g. Myers \& Benson 1983).
However, increases in instrument speed have now made it possible to survey
millimeter continuum emission over relatively large areas of the sky.
There have been two very recent results in this regard,
the first by Motte, Andr\'{e}, \& Neri (1998)
at 1.3~mm using the array bolometer on the IRAM 30~m telescope,
and the second by Testi \& Sargent (1998) at 3~mm using
the OVRO interferometer.

Motte et al. (1998) mapped the $\rho$ Ophiuchus cloud,
the closest rich cluster forming region
(see chapter by Andr\'{e} et al.)
and Testi \& Sargent (1998) mapped the Serpens
molecular cloud, a more distant, but more massive and
somewhat richer star forming region (Figure~4).
In each case, the large scale, high resolution observations reveal
a large number of embedded young protostars and also starless,
dense condensations. Both studies find that the mass spectrum
of the cores is significantly steeper, 
$x>1.1$ (where
$dN/d\ln M\propto M^{-x}$) than clump mass spectra,
$x\simeq 0.6-0.8$. The core mass spectra 
resemble the slope, $x=1.35$, 
of the stellar (Salpeter) IMF which
suggests a very direct link between cloud structure and star formation.
However, it has not yet been shown that these starless cores are
self-gravitating, which is an important step in establishing the link.

\subsection{C. The origin of the IMF}

The spectrum, lifetime, and end state of a star are primarily determined
by its mass. Consequently, the problem of understanding how the mass
of a star is determined during its formation, and the origin of the IMF,
has a very wide application in many fields from
galaxy evolution to the habitability of extrasolar planets. 
The form of the IMF is typically assumed to be invariant,
but since it is directly measureable only locally, knowing how it comes
about can help us predict how it might vary under different astrophysical
environments.

	Observations and theories of the IMF are discussed in detail
in the chapter by Adams et al.,
but here we briefly note the connection between the work
described in this chapter and the IMF.
Many explanations for the form of the IMF use as their starting point
the mass spectrum of clouds and clumps
as revealed by molecular line emission, 
$dN/d\ln M\propto M^{-x}$ with $x\simeq 0.5$.
Most such structures, however, are not
forming stars: the majority of stars form in clusters in a few of
the most massive clumps in a cloud.
An understanding of the origin
of the IMF can only come about with a more complete picture of 
the formation of star-forming clumps and the
fragmentation of these clumps down to individual star forming cores.
The unbiased continuum surveys by Motte et al. (1998) and
Testi \& Sargent (1998) are an important step in this direction.
As high resolution studies of individual cores
in cluster environments become more commonplace,
the relationship between stellar mass
and core mass can be determined.
If the core mass spectrum is indeed similar to the stellar IMF,
then the fraction of the mass of a core that goes into a star
(the star formation efficiency of the core)
is approximately independent of mass and the stellar IMF
is determined principally by the cloud fragmentation processes.
By measuring the core mass spectrum
in different clusters in a variety
of different molecular clouds, the influence of the
large scale structure and environment on the IMF can be quantified.

\mainsection{{V}.~~SUMMARY}

The study of the structure of molecular clouds is inextricably linked to
the formation of stars.
In the outer Galaxy the molecular gas is confined to spiral arms.
The inner Galaxy also shows confinement of the molecular gas to the
spiral arms, but there is some evidence for interarm molecular gas
as well.
The observed star formation rate and efficiency in the Galaxy
may be the result of only about 10\% of the molecular ISM being shielded
from the interstellar UV radiation field to an extent that matter can
drift through the magnetic field lines and condense into star forming
cores on timescales $\sim 10$~Myr.

Molecular clouds, in the outer Galaxy at least, probably result from
the compression of atomic gas entering a spiral arm. Thus, the density
inhomogeneities in clouds may simply reflect the initial non-uniform
conditions of the atomic ISM and need not be related to star formation.
At moderate densities, $\nhtwo\simlt 10^3$~\cc, cloud structures
are self-similar up to a scale set by self-gravity
and clump mass-spectra have similar power law indices
independent of the star-forming nature of the cloud.

As linewidths approach their thermal value, structures depart from
the same self-similar description. This departure may mark the boundary
between cloud evolution and star formation. Clusters of individual star
forming cores, with a mass spectrum that approaches the Salpeter IMF,
are observed in the $\rho$ Ophiuchus and Serpens clouds.
In the coming years, we can expect there to be increased observational
and theoretical effort to understand the structure, dynamics, and
distribution of these cores in a variety of star forming environments which
should lead to a better understanding of the 
relationship between the structure and evolution 
of molecular clouds and the initial mass function of stars.

\vskip 0.2in
We enjoyed many informative discussions with both
Edith Falgarone and Dick Crutcher, and we
thank them also for a thorough reading of the manuscript.
We are grateful to Mark Heyer for providing figure 1 and for a
series of interesting conversations over the last several years.
Finally, thanks to Leonardo Testi for figure 4 in advance of publication.

\vfill\eject
\null

\vskip .5in
\centerline{\bf REFERENCES}
\vskip .25in

\ref{Abergel, A., Boulanger, F., Mizuno, A., \& Fukui, Y. 1994.
    Comparative analysis of the far-infrared and (13)CO (J = 0-1)
    emissions of the Taurus complex.
    {\refit Astrophys.\ J.\ Lett.\/} 423:L59--L62.}

\ref{Adams, F.C. 1992.
    A topological/geometrical approach to the study of astrophysical maps.
    {\refit Astrophys.\ J.\/} 387:572--590.}

\ref{Bally, J., Stark, A.A., Wilson, R.W., \& Langer, W.D. 1987.
    Filamentary structure in the Orion molecular cloud.
    {\refit Astrophys.\ J.\ Lett.\/} 312:L45--L49.}

\ref{Bazell, D., \& D\'{e}sert, F.X. 1988.
    Fractal structure of interstellar cirrus.
    {\refit Astrophys.\ J.\/} 333:353--358.}

\ref{Benson, P.J., \& Myers, P.C. 1989.
    A survey for dense cores in dark clouds.
    {\refit Astrophys.\ J.\ Suppl.\/} 71:89--108.}

\ref{Bertoldi, F., \& McKee, C.F. 1992.
    Pressure-confined clumps in magnetized molecular clouds.
    {\refit Astrophys.\ J.\/} 395:140--157.}

\ref{Bertoldi, F., \& McKee, C.F. 1996.
     Self-regulated star formation in molecular clouds.
     In {\refit Amazing Light,
     A Volume Dedicated to Charles Hard Townes on His 80th Birthday},
     ed. R.Y. Chiao (New York: Springer), pp. 41--44.}

\ref{Blitz, L. 1990. The Evolution of Galactic Giant Molecular
    Clouds.  In {\refit The Evolution of the Interstellar Medium},  
    ed. L. Blitz, (ASP Press: San Francisco), pp.273--289.}

\ref{Blitz, L. 1991. Star Forming Giant Molecular Clouds.
    In {\refit The Physics of Star Formation and Early Stellar
    Evolution}, ed. C. J. Lada \& N. D. Kylafis (Dordrecht: Kluwer),
    pp 3-33.}

\ref{Blitz, L. 1993. Giant molecular clouds.
     In {\refit Protostars and Planets III},
     eds. E.H. Levy and J.I. Lunine, (Tucson: Univ.\ of Arizona Press),
     pp. 125--161.}

\ref{Blitz, L. Magnani, L. and Mundy, L. 1984.
    High-Latitude Molecular Clouds.
    {\refit Astrophys.\ J.\ Lett.\/} 282:L9--L12.}

\ref{Blitz, L., \& Shu, F.H. 1980.
    The origin and lifetime of giant molecular cloud complexes.
    {\refit Astrophys.\ J.\/} 238:148--157.}

\ref{Blitz, L., \& Stark, A.A. 1986.
    Detection of clump and interclump gas in the
    Rosette molecular cloud complex.
    {\refit Astrophys.\ J.\ Lett.\/} 300:L89--L93.}

\ref{Blitz, L., \& Thaddeus, P. 1980.
    Giant molecular complexes and OB associations.
    I: The Rosette Molecular Complex.
    {\refit Astrophys.\ J.\/} 241:676--696.}

\ref{Blitz, L., \& Williams, J.P. 1997.
    Molecular clouds are not fractal: A characteristic size scale in Taurus.
    {\refit Astrophys.\ J.\ Lett.\/} 488:L145--L149.}

\ref{Bonnell, I.A., Bate, M.R., Clarke, C.J., \& Pringle, J.E. 1997.
    Accretion and the stellar mass spectrum in small clusters.
    {\refit MNRAS\/} 285:201--208.}

\ref{Boulares, A., \& Cox, D.P. 1990.
    Galactic hydrostatic equilibrium with magnetic tension
    and cosmic-ray diffusion.
    {\refit Astrophys.\ J.\/} 365:544--558.}

\ref{Carpenter, J.M., Snell, R.L., \& Schloerb, F.P. 1995a.
    Anatomy of the Gemini OB1 molecular cloud complex.
    {\refit Astrophys.\ J.\/} 445:246--268.}

\ref{Carpenter, J.M., Snell, R.L., \& Schloerb, F.P. 1995b.
    Star formation in the Gemini OB1 molecular cloud complex.
    {\refit Astrophys.\ J.\/} 450:201--216.}

\ref{Clemens, D.P., \& Barvainis, R. 1988.
     A catalog of small, optically selected molecular clouds:
     Optical, infrared, and millimeter properties
    {\refit Astrophys.\ J.\ Suppl.\/} 68:257--286.}

\ref{Cohen, R.S., Cong, H.-I., Dame, T.M., \& Thaddeus, P.
    1980, Molecular Clouds and Galactic Spiral Structure.
    {\refit Astrophys.\ J.\ Lett.\/} 239:L53--L56.}

\ref{Cox, P., Deharveng, L. \& Leene, A. 1990.
    IRAS observations of the Rosette nebula complex.
    {\refit Astron.\ Astrophys.\/} 230:181--192.}

\ref{Crutcher, R.M. 1999.
    Magnetic fields in molecular clouds--observations confront theory.
    {\refit Astrophys.\ J.\/}, submitted.}

\ref{Crutcher, R.M., Mouschovias, T.M., Troland, T.H.,
    \& Ciolek, G.E. 1994.
    Structure and evolution of magnetically supported molecular clouds:
    Evidence for ambipolar diffusion in the Barnard 1 cloud.
    {\refit Astrophys.\ J.\/} 427:839--847.}

\ref{Dame, T.M. et al. 1987.
    A composite CO survey of the entire Milky Way.
    {\refit Astrophys.\ J.\/} 322:706--720.}

\ref{Dame, T.M., Elmegreen, B.G., Cohen, R.S., \& Thaddeus, P. 1986.
    The largest molecular cloud complexes in the first galactic quadrant.
    {\refit Astrophys.\ J.\/} 305:892--908.}

\ref{Digel, S.W., Lyder, D.A., Philbrick, A.J., Puche, D., \& Thaddeus, P.
    1996. A Large-Scale CO Survey toward W3, W4, and W5.
    {\refit Astrophys.\ J.\/} 458:561--575.}

\ref{Elmegreen, B.G. 1997a.
    Intercloud structure in a turbulent fractal interstellar medium.
    {\refit Astrophys.\ J.\/} 477:196--203.}

\ref{Elmegreen, B.G. 1997b.
    The initial stellar mass function from random sampling
    in a Turbulent Fractal Cloud.
    {\refit Astrophys.\ J.\/} 486:944--954.}

\ref{Elmegreen, B.G. 1985.
     Molecular Clouds and Star formation: An Overview
     In {\refit Protostars and Planets II},
     eds. D.C. Black \& M.S. Matthews, (Tucson: Univ.\ of Arizona Press),
     pp. 97--161.}

\ref{Elmegreen, B.G. 1993a.
    Formation of interstellar clouds and structure
    In {\refit Protostars and Planets III},
    eds. E.H. Levy and J.I. Lunine, (Tucson: Univ.\ of Arizona Press),
    pp. 97--161.}

\ref{Elmegreen, B.G. 1993b.
    The H to H$_2$ transition in galaxies: Totally molecular galaxies.
    {\refit Astrophys.\ J.\/} 411:170--177.}

\ref{Elmegreen, B.G., \& Lada, C.J. 1977.
    Sequential formation of subgroups in OB associations.
    {\refit Astrophys.\ J.\/} 214:725--741.}

\ref{Elmegreen, B.G., \& Falgarone, E. 1996.
    A fractal origin for the mass spectrum of interstellar clouds.
    {\refit Astrophys.\ J.\/} 471:816--821.}

\ref{Falgarone, E., Panis, J.-F., Heithausen, A., P\'{e}rault, M.,
    Stutzki, J., Puget, J.-L., \& Bensch, F. 1998.
    The IRAM key-project: Small-scale structure of
    pre-star-forming regions. I. Observational results.
    {\refit Astron.\ Astrophys.\/} 331:669--696.}

\ref{Falgarone, E., \& Phillips, T.G. 1990.
    A signature of the intermittency of interstellar turbulence:
    The wings of molecular line profiles.
    {\refit Astrophys.\ J.\/} 359:344--354.}

\ref{Falgarone, E., Phillips, T.G., \& Walker C.K. 1991.
    The edges of molecular clouds: Fractal boundaries and density structure.
    {\refit Astrophys.\ J.\/} 378:186--201.}

\ref{Fiege, J.D., \& Pudritz, R.E. 1998.
    Helical Fields and Filamentary Molecular Clouds.
    {\refit Mon.\ Not.\ Roy.\ Astron.\ Soc.\/} submitted (astro-ph/9901096).}

\ref{Goodman, A.A., Barranco, J.A., Wilner, D.J., \& Heyer, M.H. 1998.
    Coherence in dense cores. II. The transition to coherence.
    {\refit Astrophys.\ J.\/} 504:223--246.}

\ref{Heyer, M.H., Brunt, C., Snell, R. L., Howe, J. E., Schloerb, F. P.,
     \& Carpenter, J. M. 1998.
    The Five College Radio Astronomy Observatory CO survey
    of the outer Galaxy.
    {\refit Astrophys.\ J.\ Suppl.\/} 115:241--258.}

\ref{Heyer, M.H., Carpenter, J.M., \& Ladd, E.F. 1996.
    Giant molecular cloud complexes with optical HII regions:
    $^{12}$CO and $^{13}$CO observations and global cloud properties.
    {\refit Astrophys.\ J.\/} 463:630--641.}

\ref{Heyer, M.H., \& Schloerb, F.P. 1997.
    Application of principal component analysis to large-scale spectral
    line imaging studies of the interstellar medium.
    {\refit Astrophys.\ J.\/} 475:173--187.}

\ref{Heyer, M.H., \& Terebey, S. 1998.
    The anatomy of the Perseus spiral arm:
    $^{12}$CO and IRAS imaging observations of the W3--W4--W5 cloud complex.
    {\refit Astrophys.\ J.\/} 502:265--277.}

\ref{Heithausen, A., Bensch, F., Stutzki, J., Falgarone, E., \& Panis, J.-F.
    1998. The IRAM key project: Small-scale structure of pre-star
    forming regions. Combined mass spectra and scaling laws.
    {\refit Astron.\ Astrophys.\ Lett.\/} 331:L65--L68.}

\ref{Hester, J.J., et al. 1996.
    Hubble Space Telescope WFPC2 Imaging of M16:
    Photoevaporation and emerging young stellar objects.
    {\refit Astron.\ J.\/} 111:2349--2360.}

\ref{Heiles, C. 1996.
    in {\refit Polarimetry of the Interstellar Medium},
    eds. W.G. Roberge and D.C.B. Whittet (ASP: San Francisco).}

\ref{Heiles, C., Goodman, A.A., McKee, C.F., \& Zweibel, E.G. 1993.
     Magnetic fields in star-forming regions: Observations.
     In {\refit Protostars and Planets III},
     eds. E.H. Levy and J.I. Lunine, (Tucson: Univ.\ of Arizona Press),
     pp. 279--326.}

\ref{Houlahan, P., \& Scalo, J. 1992.
    Recognition and characterization of hierarchical interstellar structure.
    II: Structure tree statistics.
    {\refit Astrophys.\ J.\/} 393:172--187.}

\ref{Jackson, J.D. 1975. Classical Electrodynamics (Wiley: New York)}

\ref{Kramer, C., Stutzki, J., R\"{o}hrig, R., Corneliussen, U. 1998a.
    Clump mass spectra of molecular clouds.
    {\refit Astron.\ Astrophys.\/} 329:249--264.}

\ref{Kramer, C., Alves, J., Lada, C., Lada, E., Sievers, A., Ungerechts, H.,
    \& Walmsley, M. 1998b.
    The millimeter wavelength emissivity in IC5146.
    {\refit Astron.\ Astrophys.\ Lett.\/} 329:L33--L36.}

\ref{Kuchar, T.A., \& Bania, T.M. 1993.
    A high-resolution HI survey of the Rosette nebula.
    {\refit Astrophys.\ J.\/} 414:664--671.}

\ref{Kutner, M.L., Tucker, K.D., Chin, G., \& Thaddeus, P. 1977.
    The molecular complexes in Orion.
    {\refit Astrophys.\ J.\/} 215:521--528.}

\ref{Lada, C.J., Lada, E.A., Clemens, D.P., \& Bally, J. 1994.
    Dust extinction and molecular gas in the dark cloud IC5146.
    {\refit Astrophys.\ J.\/} 429:694--709.}

\ref{Lada, E.A. Strom, K.M., \& Myers, P.C. 1993.
     Environments of star formation: Relationship between molecular clouds,
     dense cores and young stars.
     In {\refit Protostars and Planets III},
     eds. E.H. Levy and J.I. Lunine, (Tucson: Univ.\ of Arizona Press),
     pp. 247--277.}

\ref{Langer, W.D., Wilson, R.W., Anderson, C.H. 1993.
    Hierarchical structure analysis of interstellar clouds
    using nonorthogonal wavelets.
    {\refit Astrophys.\ J.\ Lett.\/} 408:L45--L48.}

\ref{Larson, R.B. 1995.
    Star formation in groups.
    {\refit Mon.\ Not.\ Roy.\ Astron.\ Soc.\/} 272:213--220.}

\ref{Larson, R.B. 1985.
    Cloud fragmentation and stellar masses.
    {\refit Mon.\ Not.\ Roy.\ Astron.\ Soc.\/} 214:379--398.}

\ref{Larson, R.B. 1981.
    Turbulence and star formation in molecular clouds.
    {\refit Mon.\ Not.\ Roy.\ Astron.\ Soc.\/} 194:809--826.}

\ref{Li, W., Evans, N.J., \& Lada, E.A. 1997.
    Looking for distributed star formation in L1630: A near-infrared
    (J, H, K) survey.
    {\refit Astrophys.\ J.\/} 488:277--285.}

\ref{Lis, D.C., Pety, J., Phillips, T.G., \& Falgarone, E. 1996.
    Statistical properties of line centroid velocities
    and centroid velocity increments in compressible turbulence.
    {\refit Astrophys.\ J.\/} 463:623--629.}

\ref{Loinard, L., Dame, T.M., Koper, E., Lequeux, J., Thaddeus, P.,
     \& Young, J.S. 1996.
    Molecular spiral arms in M31.
    {\refit Astrophys.\ J.\ Lett.\/} 469:L101--L104.}

\ref{Loren, R.B. 1989.
    The cobwebs of Ophiuchus.
    I: Strands of $^{13}$CO -- The mass distribution.
    {\refit Astrophys.\ J.\/} 338:902--924.}

\ref{Maddalena, R., \& Thaddeus, P., 1985.
    A large, cold, and unusual molecular cloud in Monoceros.
    {\refit Astrophys.\ J.\/} 294:231--237.}

\ref{Magnani, L., Caillault, J.-P., Buchalter, A., \& Beichman, C.A.
     1995. A search for T Tauri stars in high-latitude molecular clouds.
     II: The IRAS Faint Source Survey catalog.
    {\refit Astrophys.\ J.\ Suppl.\/} 96:159--173.}

\ref{Mandelbrot, B.B. 1982. The Fractal Geometry of Nature
     (San Francisco: Freeman)}

\ref{Martin, H.M., Sanders, D.B., \& Hills, R.E. 1984.
    CO emission from fragmentary molecular clouds:
    A model applied to observations of M17 SW.
    {\refit Mon.\ Not.\ Roy.\ Astron.\ Soc.\/} 208:35--55.}

\ref{McLaughlin, D.E., and Pudritz, R.E. 1996.
    A Model for the Internal Structure of Molecular Cloud Cores.
    {\refit Astrophys.\ J.\/} 469:194--208.}

\ref{McKee, C.F. 1989.
    Photoionization-regulated star formation and the
    structure of molecular clouds.
    {\refit Astrophys.\ J.\/} 345:782--801.}

\ref{McKee, C.F., Zweibel, E.G., Goodman, A.A., \& Heiles, C. 1993.
     Magnetic fieldsin star-forming regions: Theory.
     In {\refit Protostars and Planets III},
     eds. E.H. Levy and J.I. Lunine, (Tucson: Univ.\ of Arizona Press),
     pp. 327--366.}

\ref{Miesch, M.S., \& Scalo, J.M. 1994.
    Statistical analysis of turbulence in molecular clouds.
    {\refit Astrophys.\ J.\/} 429:645--671.}

\ref{Mizuno, A., Onishi, T., Yonekura, Y., Nagahama, T.,
    Ogawa, H., \& Fukui, Y. 1995.
    Overall distribution of dense molecular gas and star formation
    in the the Taurus cloud complex.
    {\refit Astrophys.\ J.\ Lett.\/} 445:L161--L165.}

\ref{Moriarty-Schieven, G.H., Andersson, B.-G., \& Wannier, P.G. 1997.
    The L1457 Molecular/Atomic cloud complex: HI and CO Maps.
    {\refit Astrophys.\ J.\/} 475:642--660.}

\ref{Motte, F., Andr\'{e}, Ph., \& Neri, R. 1998.
    The initial conditions of star formation in the $\rho$ Ophiuchi main cloud:
    wide-field millimeter continuum mapping.
    {\refit Astron.\ Astrophys.\/} 336:150--172.}

\ref{Myers, P.C. 1983.
    Dense cores in dark clouds: III. Subsonic turbulence.
    {\refit Astrophys.\ J.\/} 270:105--118.}

\ref{Myers, P.C., \& Benson P.J. 1983.
    Dense cores in dark clouds. II: NH$_3$ observations and star formation.
    {\refit Astrophys.\ J.\/} 266:309--320.}

\ref{Myers, P.C., \& Goodman, A.A. 1988.
    Evidence for Magnetic and Virial Equilibrium in Molecular Clouds.
    {\refit Astrophys.\ J.\/} 326:L27--30}

\ref{Nagahama, T., Mizuno, A., Ogawa, H., \& Fukui, Y. 1998.
    A spatially complete $^{13}$CO $J=1-0$ survey of the Orion A cloud.
    {\refit Astron.\ J.\/} 116:336--348.}

\ref{Nakano, T. 1998.
    Star formation in magnetic clouds.
    {\refit Astrophys.\ J.\/} 494:587--604.}

\ref{Nakano, T., \& Nakamura, T. 1978.
    Gravitational instability of magnetized gaseous disks.
    {\refit Publ.\ Astron.\ Soc.\ Japan} 30:681--679.}

\ref{Normandeau, M., Taylor, A.R., \& Dewdney, P.E. 1997.
    The Dominion Astrophysical Observatory Galactic plane survey
    pilot project: The W3/W4/W5/HB 3 region.
    {\refit Astrophys.\ J.\ Suppl.\/} 108:279--299.}

\ref{O'Dell, C.R., \& Wong, S.K. 1996.
    Hubble Space Telescope mapping of the Orion nebula.
    I. A survey of stars and compact objects.
    {\refit Astron.\ J.\/} 111:846--855.}

\ref{Pak, S., Jaffe, D.T., van Dishoeck, E.F., Johansson, L.E.B.,
    \& Booth, R.S. 1998.
    Molecular cloud structure in the Magellanic Clouds: Effect of metallicity.
    {\refit Astrophys.\ J.\/} 498:735--756.}

\ref{Patel, N., Goldsmith, P.F., Snell, R.L., Hezel, T., \& Taoling, X. 1995.
    The large-scale structure, kinematics, and evolution of IC 1396.
    {\refit Astrophys.\ J.\/} 447:721--741.}

\ref{Sanders, D.B., Clemens, D.P., Scoville, N.Z., \& Solomon, P.M.
    1985. Massachusetts-Stony Brook Galactic plane CO survey.
    I: (b,V) maps of the first Galactic quadrant.
    {\refit Astrophys.\ J.\ Suppl.\/} 60:1--303.}

\ref{Sanders, D.B., Scoville, N.Z., \& Solomon, P.M. 1985. Giant
    Molecular Clouds inthe Galaxy II: Characteristics of Discrete
    Features. {\refit Astrophys.\ J. 289:373--387.} 

\ref{Scalo, J. 1990.
    Perception of interstellar structure: Facing complexity.
    In {\refit Physical Processes in Fragmentation and Star Formation},
    eds. R. Capuzzo-Dolcetta et al.
    (Dordrecht: Kluwer) pp. 151--171.}

\ref{Schneider, N., Stutzki, J., Winnewisser, G., \& Blitz, L. 1996.
    The nature of the molecular line wing emission
    in the Rosette molecular complex.
    {\refit Astrophys.\ J.\ Lett.\/} 468:L119--L122.}

\ref{Shu, F.H., Adams, F.C., and Lizano, S., 1987.
    Star formation in molecular clouds: Observation and theory.
    {\refit Ann.\ Rev.\ Astron.\ Astrophys.\/} 25:23--81.}

\ref{Shu, F.H., \& Li, Z.-Y. 1997.
    Magnetic forces in an isopedic disk.
    {\refit Astrophys.\ J.\/} 475:251--259.}

\ref{Simon, M. 1997.
    Clustering of young stars in Taurus, Ophiuchus, and the Orion trapezium.
    {\refit Astrophys.\ J.\ Lett.\/} 482:L81--L84.}

\ref{Solomon, P.M., \& Rivolo, A.R. 1989.
    A face-on view of the first galactic quadrant in molecular clouds.
    {\refit Astrophys.\ J.\/} 339:919--925.}

\ref{Strom, K.M., Strom, S.E., \& Merrill, K.M. 1993.
    Infrared luminosity functions for the young stellar population
    associated with the L1641 molecular cloud.
    {\refit Astrophys.\ J.\/} 412:233--253.}

\ref{Stutzki, J., Bensch, F., Heithausen, A., Ossenkopf, V., \& Zielinsky, M.
    1998. On the fractal structure of molecular clouds.
    {\refit Astron.\ Astrophys.\/} 336:697--720.}

\ref{Stutzki, J., \& G\"{u}sten, R., 1990.
    High spatial resolution isotopic CO and CS observations of M17 SW:
    The clumpy structure of the molecular cloud core.
    {\refit Astrophys.\ J.\/} 356:513--533.}

\ref{Tauber, J.A. 1996.
    The smoothness of line profiles: a useful diagnostic of clump properties.
    {\refit Astron.\ Astrophys.\/} 315:591--602.}

\ref{Testi, L., \& Sargent, A.I. 1998.
    The OVRO 3~mm continuum survey for compact sources in the Serpens core.
    {\refit Astrophys.\ J.\ Lett.\/} 508:L91--L94.}

\ref{Tomisaka, K. 1991.
    The equilibria and evolutions of magnetized, rotating, isothermal clouds.
    V: The effect of the toroidal field.
    {\refit Astrophys.\ J.\/} 376:190--198.}

\ref{van Dishoeck, E.F., \& Black, J.H. 1988.
    The photodissociation and chemistry of interstellar CO.
    {\refit Astrophys.\ J.\/} 334:771--802.}

\ref{Vazquez-Semadeni, E. 1998.
    Turbulence as an organizing agent in the ISM.
    In {\refit Interstellar Turbulence, Proceedings of the 2nd
    Guillermo Haro Conference}, eds. J. Franco and A. Carraminana,
    (Cambridge University Press), in press.}

\ref{Williams, J.P. 1998.
    The structure of molecular clouds: Are they fractal?
    In {\refit Interstellar Turbulence, Proceedings of the 2nd
    Guillermo Haro Conference}, eds. J. Franco and A. Carraminana,
    (Cambridge University Press), in press.}

\ref{Williams, J.P., \& Blitz, L. 1998.
    A multitransition CO and CS(2--1) comparison of a star-forming and a
    non--star-forming giant molecular cloud.
    {\refit Astrophys.\ J.\/} 494:657--673.}

\ref{Williams, J.P., Blitz, L., \& Stark, A.A. 1995.
    The density structure in the Rosette molecular cloud:
    Signposts of evolution.
    {\refit Astrophys.\ J.\/} 451:252--274.}

\ref{Williams, J.P., de Geus, E.J., \& Blitz, L. 1994.
    Determining structure in molecular clouds.
    {\refit Astrophys.\ J.\/} 428:693--712.}

\ref{Williams, J.P., \& Maddalena, R.J. 1996.
    A large photodissociation region around the cold, unusual cloud G216-2.5.
    {\refit Astrophys.\ J.\/} 464:247--255.}

\ref{Williams, J.P., \& McKee, C.F. 1997.
    The Galactic distribution of OB associations in molecular clouds.
    {\refit Astrophys.\ J.\/} 476:166--183.}

\ref{Wiseman, J.J., \& Adams, F.C. 1994.
    A quantitative analysis of IRAS maps of molecular clouds.
    {\refit Astrophys.\ J.\/} 435:708--721.}

\ref{Wolfire, M.G., Hollenbach, D., \& Tielens, A.G.G.M. 1993.
    CO(J=1--0) line emission from giant molecular clouds.
    {\refit Astrophys.\ J.\/} 402:195--215.}

\ref{Wood, D.O.S., Myers, P.C., \& Daugherty, D.A. 1994.
    IRAS images of nearby dark clouds.
    {\refit Astrophys.\ J.\ Suppl.\/} 95:457--501.}

\vfill\eject
\null
\end